\documentclass[12pt]{iopart}
\usepackage{graphicx} 
\begin{document}

\title[Jet Tomography and Fluctuating Initial Conditions]
{Sensitivity of Azimuthal Jet Tomography to Early Time Energy-Loss 
at RHIC and LHC}

\author{Barbara Betz$^1$, Miklos Gyulassy$^1$, and Giorgio Torrieri$^{2}$}

\address{$^1$ Department of Physics, Columbia University, New York, 
10027, USA}
\address{$^2$ Frankfurt Institute for Advanced Studies (FIAS), 
Frankfurt am Main, Germany}
\ead{betz@phys.columbia.edu}
\begin{abstract}

We compute the path-length dependence of energy-loss for 
higher azimuthal harmonics of jet-fragments in a generalized 
model of energy-loss that can interpolate between pQCD and AdS/CFT 
limits and compare results with Glauber and CGC/KLN initial 
conditions. We find, however, that even the high-$p_T$ second moment is
most sensitive to the poorly known early-time evolution
during the first fm/c. Moreover, we demonstrate that quite generally
the energy and density-dependence leads to an overquenching of high-$p_T$
particles relative to the first LHC $R_{AA}$-data, once the parameters of the 
energy-loss model are fixed from $R_{AA}$-data at RHIC.


\end{abstract}


\section{Introduction}
Heavy-ion collisions at the Relativistc Heavy Ion Collider (RHIC)
indicate the production of an opaque (i.e.\ strongly jet-suppressing)
\cite{whitepapers}, fast-thermalizing medium. However, so far neither 
the initial conditions of the collisions nor the microscopic dynamics 
of the jet-energy loss are conclusively understood. 

To characterize the initial conditions, one usually uses either
the Glauber model, describing incoherent superpositions of proton-proton
collisions, or the ``Color Glass Condensate'' (CGC), given e.g.\ by the 
KLN model, where saturation effects are taken into account \cite{initial}.
On the other hand, the jet-energy loss can either be described as multiple 
scatterings of the hard parton \cite{pQCD}, specific of a weakly-coupled pQCD 
medium, or using the AdS/CFT correspondence where the problem of a 
parton stopped in a thermal medium is related to the 
problem of a string falling into a $5$-dimensional black hole 
\cite{AdSCFT}.

Focussing on the different path-length dependences of $dE/dx\sim l$ 
for pQCD \cite{pQCD} (as it occurs in the presence of coherence effects
like in the high-density LPM limit) and $dE/dx\sim l^2$ for AdS/CFT calculations \cite{AdSCFT}, the first
simple jet absorption model that simultaneously describes the $R_{AA}(N_{part})$
and the $v_2(N_{part})$ at RHIC energies for high-$p_T$ particles was given in
Refs.\ \cite{Jia}. It showed that after fixing the coupling such that
the most central data point for $R_{AA}$ is reproduced, the 
$R_{AA}(N_{part})$ can be described for both pQCD and AdS/CFT-like
energy loss. However, in case of a pQCD-like energy loss, the $v_2(N_{part})$ 
is underpredicted for both Glauber and CGC initial conditions, while 
for an AdS/CFT-like energy loss and CGC initial conditions, the
$v_2(N_{part})$ can be well described.

Here we want to examine if a generic energy-loss ansatz
that includes both a path-length and an energy dependence  
confirms the above conclusion that only CGC/KLN initial conditions
and an AdS/CFT energy-loss can describe both the $R_{AA}(N_{part})$
and the $v_2(N_{part})$ appropriately.
\begin{figure}[t]
\centering
  \includegraphics[scale = 0.53]{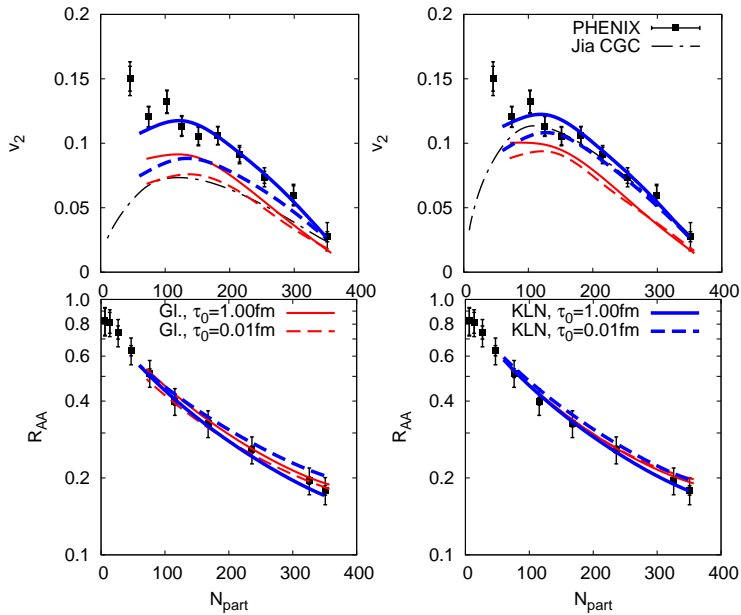}
  \caption{$R_{AA}$ and $v_{2}$ as a function of $N_{part}$ 
           at a $p_T=10$~GeV and RHIC energies for Glauber
           (red lines) and KLN (blue lines) initial conditions. The  
           initialization time is either $\tau_0=1$~fm (solid lines) or
           $\tau_0=0.01$~fm (dashed lines). The data and the black
           dashed-dotted line are taken from Refs.\ \cite{Jia}.}
  \label{Fig1}
\end{figure}

\section{Jet Tomography}
Since all dependences on the intrinsic scales of the system 
($T_c, \Lambda_{QCD}$, etc.) disappear in the high-temperature limit, a 
generic energy-loss rate $dE/dx$ is given by an arbitrary combination 
of dimensionful parameters constrained by the total dimension of the 
observable and the requirement that faster particles and hotter media 
result in a bigger suppression. We choose \cite{us}
\begin{eqnarray}
\hspace*{-0.4cm}
\frac{dE}{dx}(\vec{x}_0,\phi,\tau)&=&
-\kappa P^a\tau^{z} T^{z-a+2}[\vec{x}_0+\hat{n}(\phi)\tau,\tau],
\label{GenericEloss}
\end{eqnarray}
where $\kappa$ is the coupling, $P$ is the momentum of the jet(s) considered and $a,z$ are parameters 
controlling the jet energy (momentum) and path-length dependence, respectively. 
In the Bethe-Heitler limit $a=1$ and $z=0$, while in the deep LPM pQCD 
limit $a\sim0$ and $z\sim 1$. If $a=0$ and $z=2$, our model coincides 
with the model referred to as "AdS/CFT" in Refs.\ \cite{Jia}. 
However, on-shell AdS/CFT calculations \cite{AdSCFT} show that $a=1/3$ 
and $z=2$, thus we are going to consider $a=1/3$ throughout
the whole paper. In a static medium, $dE/dx \sim \tau^z$, while in a 
dynamic medium, $dE/dx$ will aquire additional powers of $\tau$ due to 
the dependence of temperature on $\tau$. Here, we assume a $1$D Bjorken expansion. 
In contrast to Refs.\ \cite{Jia}, $\kappa$ is a dimensionless parameter.
It is always fitted to reproduce the most central value for $R_{AA}$ at RHIC energies.

Choosing $\tau_0=1$~fm, in line with recent hydrodynamic calculations
\cite{heinzrecent}, we see (cf.\ Fig.\ \ref{Fig1}) that CGC/KLN initial
conditions get close to the RHIC data for both pQCD and AdS/CFT-like energy
loss, while Glauber initial conditions underpredict the data.
However, choosing a much smaller $\tau_0$ (in Refs.\ \cite{Jia}, $\tau_0=0$~fm), the $v_2(N_{part})$ is 
reduced for both pQCD and AdS/CFT-like energy loss, increasing the 
difference between the pQCD and AdS/CFT results as seen in Refs.\ 
\cite{Jia}, and raising the question of the physical meaning of $\tau_0$.

Setting $\tau_0=1$~fm means to assume that there is no energy loss 
within the first fm. PQCD does not give any excuse for this assumption and
thus $\tau_0=0$~fm would be a natural assumption. However, $\tau_0$
also describes the formation time of hydrodynamics which seems to be
$\tau_0\sim 1$~fm \cite{heinzrecent}.
On the other hand, setting $\tau_0=1$~fm is also equivalent to the 
AdS/CFT result that the energy loss is suppressed at early times. 
Please note that for AdS/CFT the $dE/dx\sim l^2 \sim \tau^2$ dependence leads
to a suppressed energy loss at early times where the suppression is larger 
than in the pQCD case with $dE/dx\sim l \sim \tau$. 

Thus, it is important to see that
the $v_2$ of high-$p_T$ particles is sensitive to short-distance properties,
suggesting that there is either weak coupling with a $\tau_0\sim 1$~fm or
strong coupling which in itself features the suppression of energy loss
at early times.

\begin{figure}[t]
\centering
  \includegraphics[scale = 0.53]{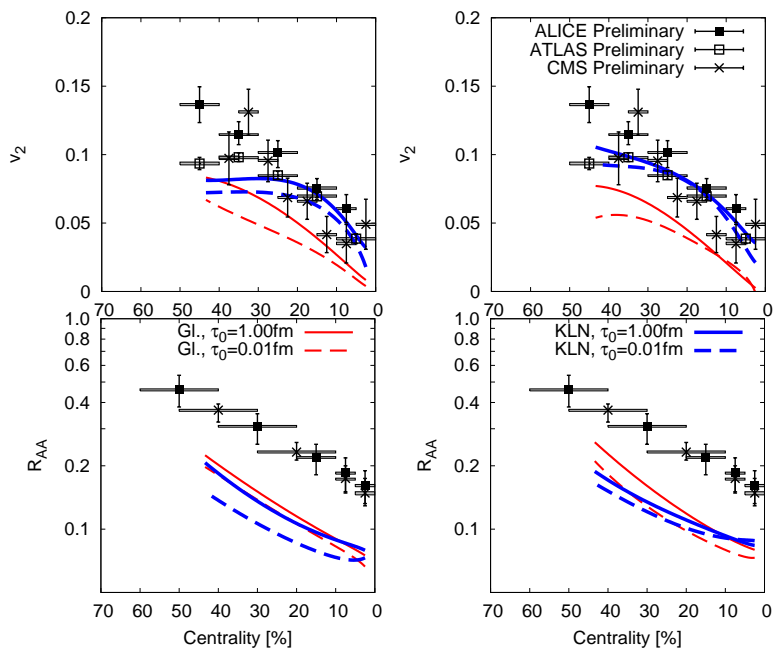}
  \caption{$R_{AA}$ and $v_{2}$ as a function of centrality
           at a $p_T=10$~GeV and LHC energies for Glauber
           (red lines) and KLN (blue lines) initial conditions. The 
           initialization time is either $\tau_0=1$~fm (solid lines) or
           $\tau_0=0.01$~fm (dashed lines). The data are taken from 
           Refs.\ \cite{DataLHC} for the $p_T$ bin just below $10$~GeV.}
  \label{Fig2}
\end{figure}    

Calculating the nuclear modification factor and elliptic flow for
pions at LHC energies while keeping the values for $a$ and $\kappa$ fixed
compared to RHIC energies, leads to an underprediction of the $R_{AA}$ as a function of centrality
as shown in Fig.\ \ref{Fig2} for both pQCD and AdS/CFT-like energy loss 
and different values of the initialization time $\tau_0$. This is a puzzle
common to all density-dependent energy-loss prescriptions 
[cf.\ Eq.\ (\ref{GenericEloss})], as discussed in Ref.\ \cite{Horowitz:2011gd}.

\section*{Acknowledgments}
B.B.\ is supported by the Alexander von Humboldt foundation via a Feodor 
Lynen fellowship. M.G.\ and B.B.\ acknowledge support from DOE under
Grant No.\ DE-FG02-93ER40764.
G.T.\ acknowledges the financial support received from the Helmholtz 
International Center for FAIR within the framework of the LOEWE program
(Landesoffensive zur Entwicklung Wissenschaftlich-\"Okonomischer
Exzellenz) launched by the State of Hesse. The authors thank A.\ Dumitru
for providing his KLN code to simulate CGC initial conditions.

\end{document}